\documentclass[aps,pre,twocolumn,superscriptaddress,showpacs]{revtex4}
\usepackage[pdftex]{graphicx}
\usepackage{amssymb,amsmath}
\usepackage{color}
\usepackage{float}
\usepackage{bm}
\newcommand{\ignore}[1]{}

\begin{document}

\title{Is the Voter Model a model for voters?}

\author{Juan Fern\'andez-Gracia}
\affiliation{IFISC (CSIC-UIB), 07122 Palma de Mallorca, Spain}
\author{Krzysztof Suchecki}
\affiliation{Faculty of Physics and Center of Excellence for Complex Systems Research, Warsaw University of Technology, Koszykowa 75, 00-662 Warsaw, Poland}
\author{Jos\'e J. Ramasco}
\affiliation{IFISC (CSIC-UIB), 07122 Palma de Mallorca, Spain}
\author{Maxi San Miguel}
\affiliation{IFISC (CSIC-UIB), 07122 Palma de Mallorca, Spain}
\author{V\'ictor M. Egu\'iluz}
\affiliation{IFISC (CSIC-UIB), 07122 Palma de Mallorca, Spain}

\date{\today}

\begin{abstract}
The voter model has been studied extensively as a paradigmatic opinion dynamics' model. However, its ability for modeling real opinion dynamics has not been addressed. We introduce a noisy voter model (accounting for social influence) with agents' recurrent mobility (as a proxy for social context), where the spatial and population diversity are taken as inputs to the model. We show that the dynamics can be described as a noisy diffusive process that contains the proper anisotropic coupling topology given by population and mobility heterogeneity. The model captures statistical features of the US presidential elections as the stationary vote-share fluctuations across counties, and the long-range spatial correlations that decay logarithmically with the distance. Furthermore, it recovers the behavior of these properties when the geographical space is coarse-grained at different scales from the county level through congressional districts and up to states. Finally, we analyze the role
of the mobility range and the randomness in decision making which are consistent with the empirical observations.
\end{abstract}


\maketitle

Opinion dynamics focuses on the way different options compete in a population, giving raise to either consensus (every individual holding the same opinion or option) or coexistence of several opinions. Many theoretical efforts have been devoted to clarify the implications on the macroscopic outcome, among other aspects, of different interaction mechanisms, different topologies of the interaction networks, the inclusion of opinion leaders or of zealots, external fields~\cite{RevModPhys.81.591,maxi2005}. To advance our understanding on social phenomena these theoretical efforts need to be complemented with empirical~\cite{Sneppen,Kandler,Abrams} and experimental results~\cite{Bond2012,Centola2010,Gallupa2011,Lorenz2011}. In this context elections offer an opportunity for contrasting opinion dynamics' models with empirical results~\cite{Fortunato2012}. On one hand, the data are publicly available in many countries, with a good level of spatial resolution and several temporal observations. On the
other hand, there is evidence that voting behavior is strongly influenced by the social context of the individuals~\cite{Centola2010,Context,social_environment,social_calculus_voting,zuckerman,Fowler2005,Bond2012,Lazarsfeld1948,rogers_diff_innov,Christakis_obesity, emotions_infectious,Rendell2010,Lorenz2011}. Thus it is natural to model electoral processes as systems of interacting agents with the aim of explaining the statistical regularities~\cite{PhysRevLett.99.138701,Chatterjee2013,Klimek2012,Borghesi_spatial1,Borghesi_spatial2,Enikolopov2012,Costa_filho1999,Costa_filho2003,Araripe2009,Bernardes2002,Lyra2003,Travieso2006,Araripe2006,Gonzalez2004,Andresen2008,Hernandez-Saldana2009,Araujo2010,Kim2003}, as for example, the universal scaling of the distribution of votes in proportional elections~\cite{PhysRevLett.99.138701,Chatterjee2013} or signatures of irregularities in the democratic process~\cite{Klimek2012,Enikolopov2012}.

In this work we asses the capacity of the voter model to capture real voter choices and propose a microscopic foundation for modeling voting behavior in elections. The model is based on social influence and recurrent mobility (SIRM): social influence will be modeled as a noisy voter model, while recurrent mobility serves as a proxy of the social context. In the voter model each agent updates its state by randomly copying the opinion of one of its neighbors~\cite{holley,vazquez_eguiluz,Suchecki2005}. We will consider that agents interact at home and at work locations according to their commuting pattern~\cite{Gonzalez2008,Song2010}. We first obtain the statistical features of the US presidential elections and then we introduce and analyze the model.

\begin{figure}
 \begin{center}
 \includegraphics[width=8.6cm]{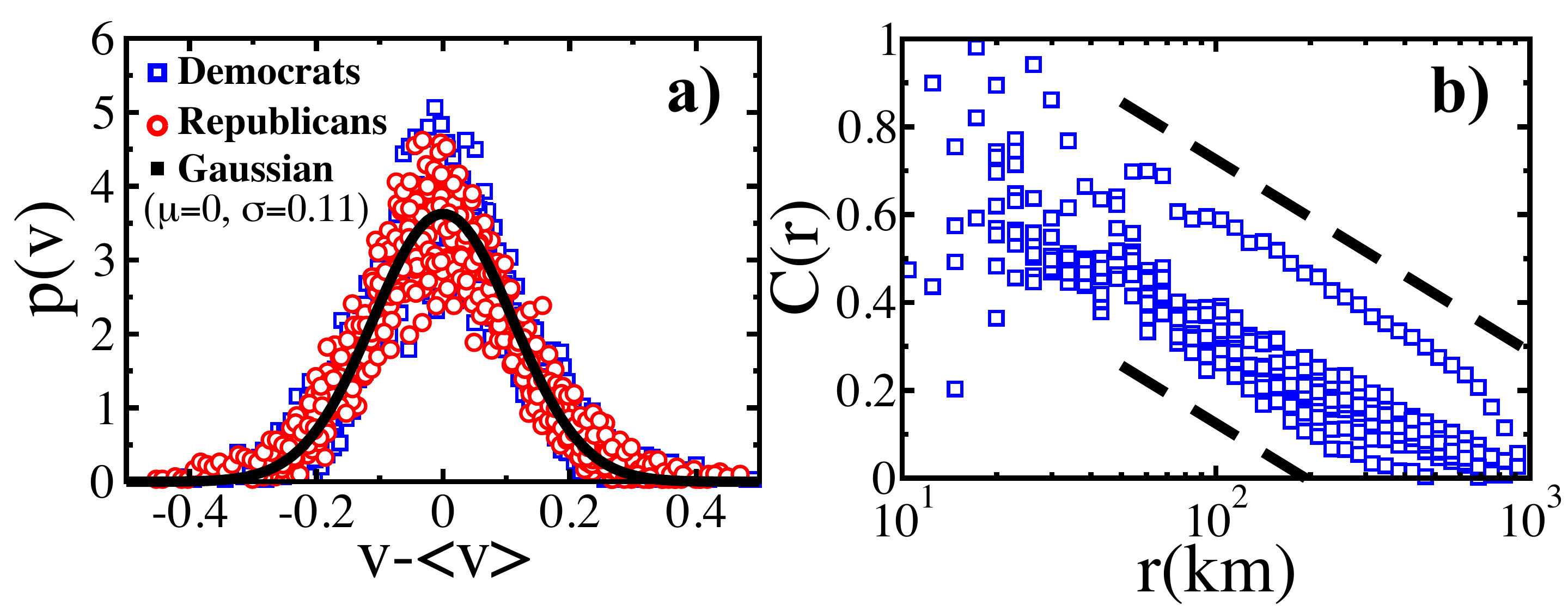}
 \caption{(Color online) US electoral results $1980$-$2012$. a) County vote-share probability density functions. b) Spatial vote-share correlations as a function of distance. The dashed lines indicate logarithmic decay. \label{elect_results}}
  \end{center}
\end{figure}

\emph{Statistical regularities in elections.-}
The analysis is focused on US presidential elections from $1980$ to $2012$~\cite{SM} due to the combination of data availability and an almost bipartisan system. The vote-share per county $v$, that is, percentage of votes in a county, for any of the two main parties is distributed following approximately a Gaussian distribution (Fig.~\ref{elect_results}a), consistently with observations in other countries~\cite{Klimek2012}. The average vote-share over all counties changes from election to election but the width remains approximately constant for each year, with a standard deviation of $\sigma_e \simeq 0.11$. We also find that the spatial correlation of the vote-shares decays logarithmically with the geographical distance (Fig.~\ref{elect_results}b), as reported
previously for turnout and winner party vote-shares~\cite{Borghesi_spatial1,Borghesi_spatial2}. The spatial correlation function is computed as
\begin{equation}
 C(r)=\frac{\langle v_{i}v_{j} \rangle|_{d(\vec{r}_i,\vec{r}_j)=r}-\langle v \rangle^2}{\sigma^2(v)},
\end{equation}
where $\langle v \rangle$ is the average vote-share over all the cells, $\sigma^2(v)$ its standard deviation, and $\langle v_{i}v_{j} \rangle|_{d(\vec{r}_j,\vec{r}_i)=r}$ is averaged over pairs of cells separated a distance $r$. The stationarity of the vote-share dispersion and the logarithmic decay of the spatial correlations will be considered as generic of the fluctuations in electoral dynamics.

\begin{figure}
\begin{center}
 \includegraphics[width=8.6cm]{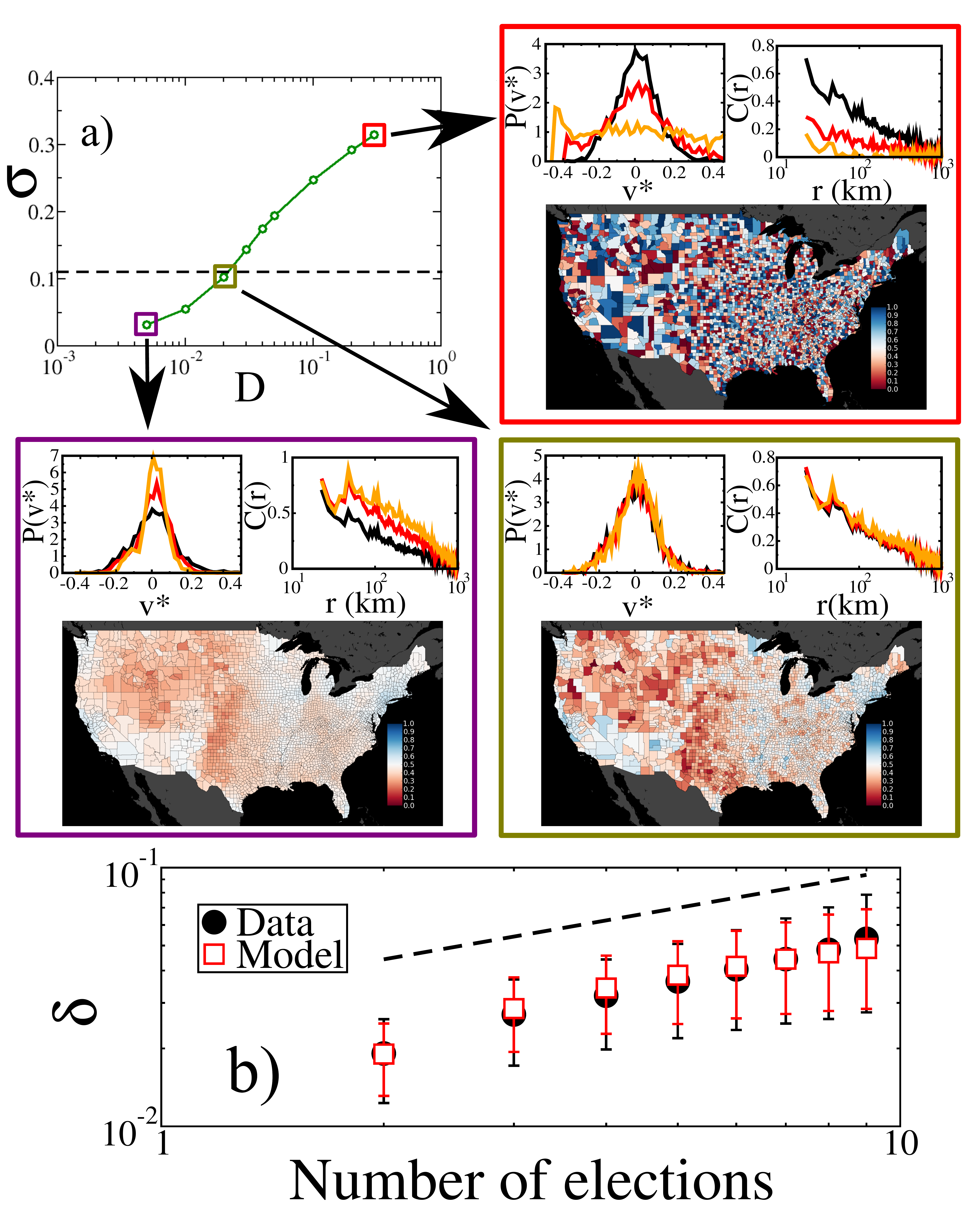}
 \caption{(Color online) a) Vote-share standard deviation versus noise intensity $D$. The dashed black line marks the dispersion of the empirical data ($\sigma_e=0.11$). Boxes surrounding the main plot display results obtained with the level of noise marked as squares and include the distribution of vote-shares shifted to have zero mean, and their spatial correlations. Black curves are initial conditions. In the red box, the red curve is for 10 MC steps, and the orange for 20 MC steps; In the green box, the times are 100 MC steps (red) and 200 MC steps (orange); in the purple box, 40 MC steps (red) and 140 MC steps (orange). b) The average dispersion in the democrat vote-share is plotted versus the number of elections. Best agreement is obtained for $2.5 \text{MCsteps/year}$.
\label{model_calibration}}
\end{center}
\end{figure}

\emph{The model.-} In the SIRM model $N$ agents live in a spatial system divided in non-overlapping cells~\cite{footnote2}. The agents are distributed among the different cells according to their residence cell. The number of residents in a particular cell $i$ is $N_i$. While many of these individuals may work at $i$, some others will work at different cells. This defines the fluxes $N_{ij}$ of residents of $i$ recurrently moving to $j$ for work. By consistency,  $N_{i}=\sum_j N_{ij}$. The working population at cell $i$ is $N'_i=\sum_j N_{ji}$ and the total population in the country is $N = \sum_{ij}\, N_{ij}$.

We describe agents' opinion by a  binary variable ($+1$ or $-1$). The number of individuals holding opinion $+1$, living in county $i$ and working at $j$ is $V_{ij}$, thus $V_i=\sum_lV_{il}$ is the number of voters living in $i$ holding opinion $+1$; $V'_j=\sum_lV_{lj}$ is the number of voters working at $j$ holding opinion $+1$. We assume that each individual interacts with people living in her own location with a probability $\alpha$, while with probability $1-\alpha$ she does so with individuals of her work place. Once an individual interacts with others, its opinion is updated following a noisy voter model \cite{Fowler2005,holley,vazquez_eguiluz,Sood2005,kinetic}: an interaction partner is chosen and the original agent copies her opinion imperfectly (with a certain probability of making mistakes). The evolution of the system can be expressed in terms of the transition rates:
\begin{align}\label{rates} &r_{ij}^-(\mathcal{V})=V_{ij}\left[\alpha\frac{N_i-V_i}{N_i}+(1-\alpha)\frac{N'_j-V'_j}{N'_j}\right]+N_{ij}\frac{D}{2}\eta_{ij}^-(t),\nonumber\\ &r_{ij}^+(\mathcal{V})=(N_{ij}-V_{ij})\left[\alpha\frac{V_i}{N_i}+(1-\alpha)\frac{V'_j}{N'_j}\right]+N_{ij}\frac{D}{2}\eta_{ij}^+(t),
\end{align}
where $\mathcal{V}=\{V_{ij}\}$ is the configuration of the system according to the set of variables $V_{ij}$, and $r_{ij}^{\pm}(\mathcal{V})$ are the rates of change of $V_{ij}$ by one unit to $V_{ij}\pm 1$. These rates include recurrent mobility and so they are different from those obtained for random diffusion processes \cite{voter_mobility}. The variables $\eta_{ij}^{\pm}(t)$ are noise terms accounting for imperfect imitation, modeled as Gaussian noise with zero mean and $\langle \eta_{ij}^a(t) \eta_{kl}^b(t')\rangle=\delta(t-t')\,\delta_{ab}\, \delta_{ik}\, \delta_{jl}$~\cite{maxi_external_noise}. At the leading order, which corresponds to taking into account only the external noise coming from imperfect imitation, while the internal noise coming from the finite number of voters is neglected, the set of stochastic differential equations for $v_{ij}=V_{ij}/N_{ij}$ is
\begin{equation}
\label{eq_evol}
 \frac{d v_{ij}}{dt} =  \, \alpha\, \sum_lA_{ijl} v_{il} + (1-\alpha)\sum_lB_{ijl} v_{lj} + D \, \eta_{ij}(t),
\end{equation}
with $A_{ijl}=\frac{N_{il}}{N_i}-\delta_{jl}$ and $B_{ijl}=\frac{N_{lj}}{N'_j}-\delta_{li}$ ~\cite{SM}. The first term on the right hand side describes interactions among agents who live in $i$ and work elsewhere, while the second term follows from the interactions among agents who work in $j$ and live elsewhere. The last term is noise coming from a combination of $\eta_{ij}^+(t)$ and $\eta_{ij}^-(t)$: $\eta_{ij}(t)$ is also a Gaussian noise with zero mean and $\langle \eta_{ij}(t) \eta_{kl}(t')\rangle=\delta(t-t')\, \delta_{ik}\, \delta_{jl}$. This term represents imperfect imitation and accounts for the combined effect of all other influences different from the interaction between peers. This includes opinion drift, local media or free
will of the individuals. When $D\neq 0$ the microscopic rules lead to a noisy diffusive equation, in agreement with previous models of mesoscopic electoral dynamics ~\cite{Borghesi_spatial1,Borghesi_spatial2}. The equation corresponds to an Edwards-Wilkinson equation on a disordered medium, described by the coupling matrices $A$ and $B$.
In the absence of imperfect imitation ($D=0$), Eq.~(\ref{eq_evol}) can be written as a Laplacian $\frac{d}{dt} \vec{v} =\mathcal{L}  \vec{v}$. This implies a homogeneous asymptotic configuration and the existence of a globally conserved variable, namely the total number of voters holding opinion $+1$, $V=\sum_{ij}V_{ij}$~\cite{SM,Klemm2012}.

When simulating the model, we integrate the stochastic process by updating the values of the number of agents holding opinion $+1$ in each cell $ij$, $V_{ij}$, using binomial distributions with the rates in Eq.~(\ref{rates}). At each Monte Carlo step we update all cells in a random order. Therefore we simulate the original master equation of the process.


\emph{Model calibration.-}
We apply the model to the US presidential elections identifying the cells with the counties. The populations and commuting fluxes $N_{ij}$ are obtained from the $2001$ census~\cite{census} and are input data for the SIRM model. This framework can be applied to any country, besides the US, or territorial division (counties, municipalities, provinces, states, etc). Besides these data, there are two free parameters: $D$ and $\alpha$. The parameter $\alpha$ provides a measure of the relative intensity and duration of the social relations at work and at home. According to the survey on time use of the Bureau of Labor Statistics~\cite{labor}, the average individual spends daily almost $8$ hours at work and the rest of time at her home location. Out of this home time, close to another $8$ hours are spent sleeping. Thus $\alpha$ will be set at $1/2$, although other values give rise to qualitatively similar results (Fig.~S10 in \cite{SM}), as long as $\alpha\neq0$ or $1$ in 
which cases the system would consist of disconnected patches and thus there would not be any spatial diffusion.

To calibrate the noise intensity $D$, the SIRM model is run for a set of values of $D$ taking as initial condition the results for the elections of the year $2000$. The system is evolved for $1000$ Monte Carlo steps and then the standard deviation $\sigma$ of the vote-share distribution is measured (panel a) in Fig.~\ref{model_calibration}).
\begin{figure}
 \begin{center}
 \includegraphics[width=8.6cm]{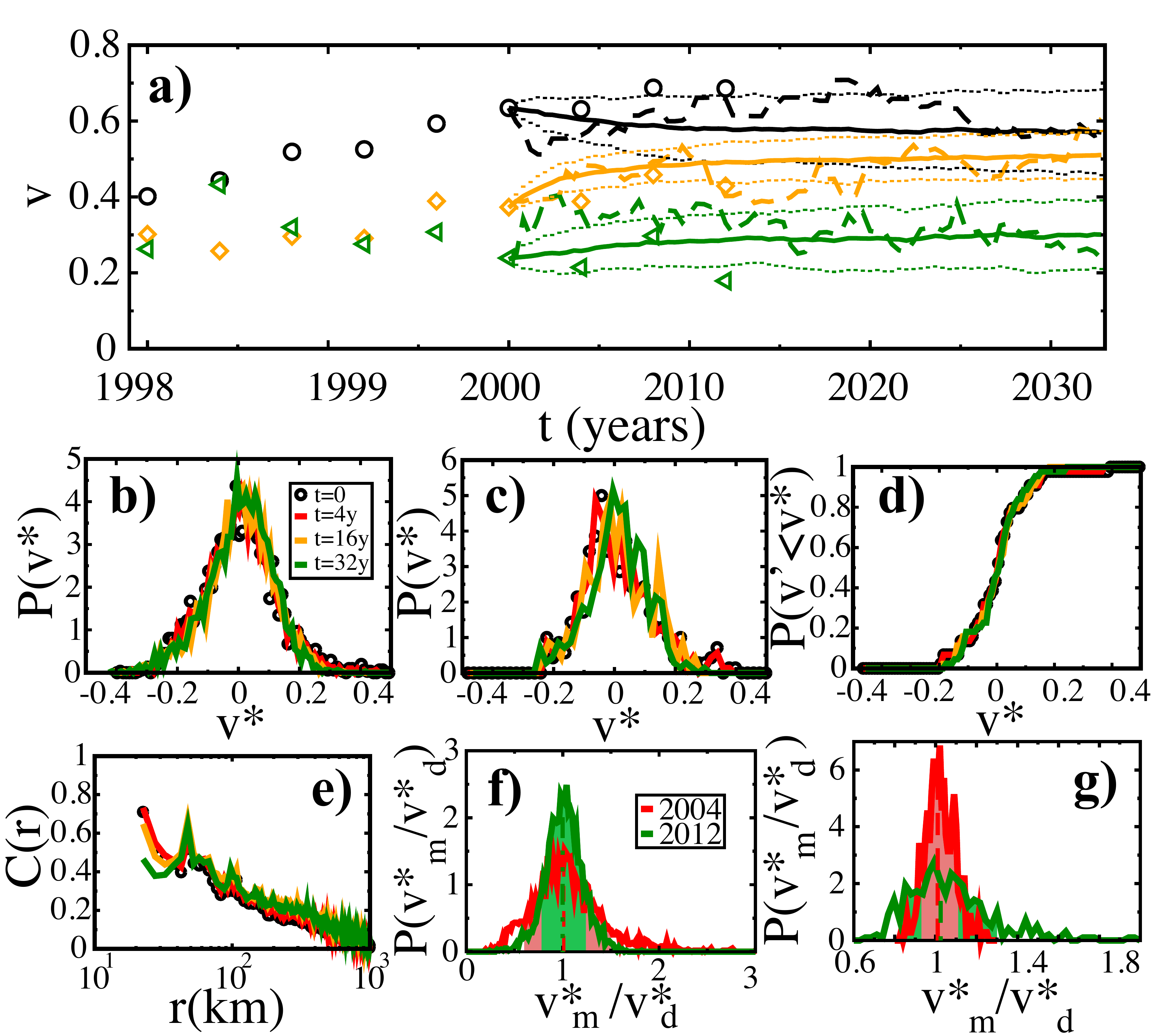}
 \caption{(Color online) Parameters of the simulation are $\alpha=1/2$, $D=0.021$. a) Time traces of the vote-shares for Democrats in different counties; one with high population, Los Angeles CA, (black symbols and curves, $9.5\times 10^6$ inhabitants); one with a medium population, Blane ID, (in orange, $19\times 10^3$ inhabitants); and one with low population, Loving TX (green line, $67$ inhabitants). Symbols represent data, dashed lines represent the results of a single realization of the model with initial conditions taken from year 2000 data, solid lines represent the average of 100 realizations of the model; dotted lines their standar deviation. b), c) and d) Democratic vote-share probability density functions (except for d), showing the cumulative pdf) as predicted by the model for counties, congressional districts and states, respectively. Initial condition at $t = 0$ (black circles): vote-shares obtained from the $2000$ 
elections. e) Vote-share spatial correlations as a function of the distance. f) and \textbf{g)} Distribution of ratio between model predictions and data observations for the
Democratic vote-shares at county level (f) and for congressional districts (g). The colored areas mark $80\%$
confidence intervals.
 \label{model_results}}
\end{center}
\end{figure}
Best agreement is obtained for $D=0.03$ which is taken as the level of noise for the simulations of the model. When the noise intensity is too low we find basically a diffusive process, where the vote share distribution narrows and the correlations grow ($D=0.005$ in Fig.~\ref{model_calibration}). In contrast, for larger $D$ the noise is dominating the results ($D=0.35$ in Fig.~\ref{model_calibration}). The vote share distribution widens as time goes by and the spatial correlations vanish. For $D=0.021$ the standard deviation of the vote-share distribution of the model has the same value as the data. Not only the standard deviation is matched, but also the shape of the vote-share distribution agrees with the empirical one. The distribution, in addition, becomes stationary in time. Furthermore, although we did not take spatial correlations into account for the calibration, they show a stationary logarithmic decay for this value of noise intensity $D$.

Finally, we set the equivalence between the Monte Carlo (MC) steps and the real time between elections (Fig.~\ref{model_calibration}b). Eq.~(\ref{eq_evol}) is written in arbitrary time units and is related to the updates by $dt=1/N$\cite{footnote3}. Sets of electoral results are produced with the model with $D = 0.021$ and with a fixed number of Monte Carlo steps between elections. Then the standard deviation $\delta$ of the vote-share trajectory for each county as a function of a the number of consecutive elections is computed. Averaging over all different counties and comparing with empirical data, we find that both curves grow as $\sqrt{n}$, where $n$ is the number of elections considered (error bars correspond to the dispersion of $\delta$ across counties), reminiscent of a random walk. Both curves have the best overlap when we set 10~MCsteps/election
(equivalently 2.5~MCsteps/year).

\emph{Results.-}
The stochasticity of the model introduces uncertainty in the temporal evolution of the vote-shares as can be appreciated for three counties in Fig.~\ref{model_results}a. Once the average value is discounted, the shape of the distribution of vote-shares is similar to the one observed in the empirical data (Fig.~\ref{model_results}b): the stationarity and the particular functional shape of the distributions are captured by the model. This occurs not only at county level (Fig.~\ref{model_results}b) but also at other coarse-grained geographical scales such as congressional districts (Fig.~\ref{model_results}c) and states (Fig.~\ref{model_results}d). This relates to the ability of the model to properly capture the spatial correlations in the data (Fig.~\ref{model_results}e) and Fig. S7 in ~\cite{SM} for a comparison with reshuffled data).

The goodness of the model is also assesed by a direct comparison between model predictions and data for vote-share fluctuations. In Fig.~\ref{model_results}f, g, we show the distribution of the ratios between model and data of the vote-shares deviations  from the national average, $v_i-\langle v \rangle$, where $\langle \cdot \rangle$ denotes spatial average (not average over realizations of the model). We evolve the model for an election, starting with the initial conditions from the electoral results from year $2000$, and compare with the electoral results from year $2004$, finding that $80\%$ of the ratios fall between $0.6$ and $1.5$. These numbers become $0.9$ and $1.1$ at the congressional district level, attesting the quality of the model predictions.

\begin{figure}
 \begin{center}
 \includegraphics[width=8.6cm]{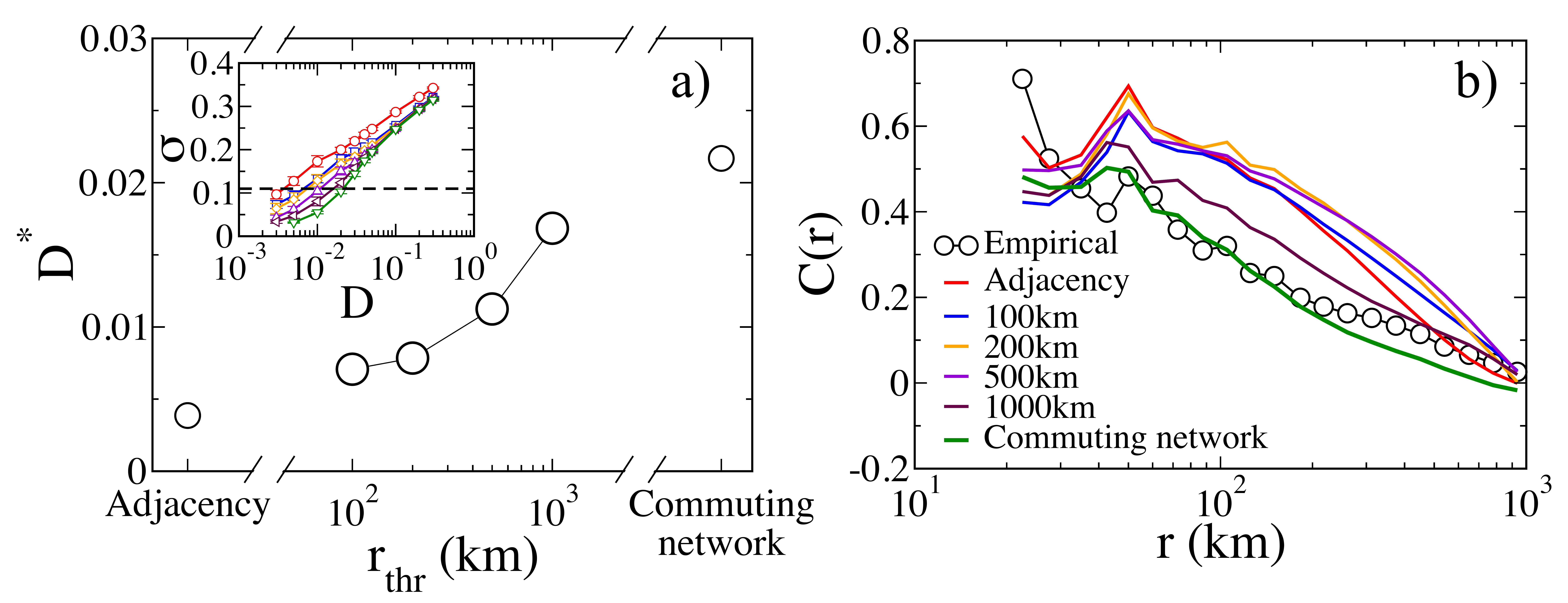}
 \caption{(Color online) a) Calibrated noise intensity $D^*$ on networks thresholded at different distances. b) Vote-share spatial correlations on different networks. \label{model_networks}}
\end{center}
\end{figure}

As a final point we investigate the role played by the mobility network on the model results. The links connecting only geographically neighboring counties can be extracted and used as a baseline network. The rest of the links are then added filtering by the distance that separates the centroid of the residence county to that of the work county. The result of performing this operation is a network that includes more and more links as the threshold of the filter is increased. The model has to be calibrated for each new network (Fig.~\ref{model_networks}a). Once the optimal value for the noise level of the imperfect imitation $D^*$ is found, the model simulations running on different networks can be compared with the empirical data. In Fig.~\ref{model_networks}b, we show how the vote-share spatial correlations change when the network is modified. Long links are important to recover correlation values similar to those observed empirically.

\emph{Discussion.-}
We have introduced a microscopic model for opinion dynamics whose main ingredients are social influence (modeled as a noisy voter model), mobility and population heterogeneity. The model can be approximated by a noisy diffusion equation on an anysotropic substrate that is given by the highly heterogeneous population and commuting data. It reproduces generic features of the vote-share fluctuations observed in data coming from three decades of presidential elections. It is important to note that the model is not aimed at predicting the winning party, only the local fluctuations over the national average vote-share. In this sense, it is able to capture the empirical distributions of vote-share fluctuations, the spatial correlations and even the evolution of the local vote-share fluctuations. This agreement between model predictions and empirical data are maintained when the geographical areas considered are coarse-grained, showing thus that the model accounts for the main mechanisms at play in the dynamics of
the system at different scales. We have studied, besides, the relevance of the mobility range for the quantitative agreement of the model. Despite the various heterogeneity sources of the system (population, geography, topology and commuter fluxes), the model still displays logarithmic spatial correlations as in a two-dimensional diffusion \cite{Ma}. This robustness is connected to the spatial decay of the coupling across cell~\cite{SM,Newman2013}. The field of random walks on heterogeneous media could also provide valuable insight~\cite{Brockmann}.

The present work offers -with the use of demographic data as input- a comparison of a theoretical model with real data, which is used both for calibration and to evaluate the results. It proposes a path for further research in opinion dynamics since it establishes a method to bridge the gap existing between microscopic mechanisms of social interchange and macroscopic results of surveys and electoral processes. One limitation of the work is the use of census data, which translates in a lack of fine structure for the interaction network. The use of digital data will provide the necessary information to fill this gap. Another important issue is the dynamics of the average vote-share. To this end further elements need to be included, as for example the effects of social and mass media.

J.F.-G. receives economic support from the Conselleria d'Educaci\'o, Cultura i Universitats of the Government of the Balearic Islands and the ESF. J.J.R. acknowledges funding from the Ram\'on y Cajal program of the Spanish Ministry of Economy (MINECO). Partial support was received from MINECO and FEDER through projects, (FIS2011-24785), (FIS2012-30634) and from the EU Commission through the FP7 projects EUNOIA and LASAGNE.

\onecolumngrid


\appendix
  
\section{Commuting data}
\begin{figure}[H]
 \centering
 \includegraphics[width=0.9\textwidth]{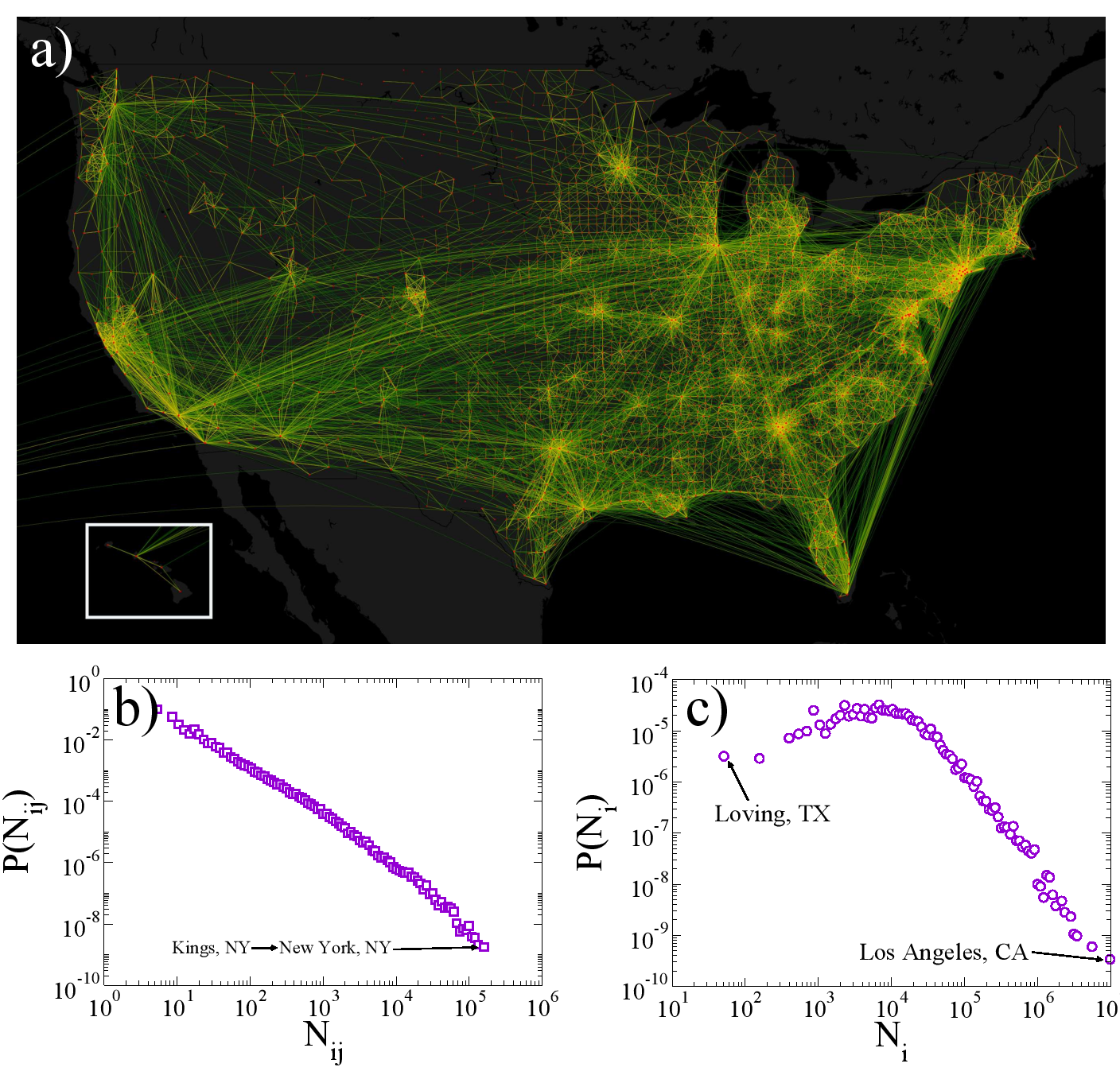}
 \caption{\textbf{Commuting data.} \textbf{a)} Map showing $10\%$ of all commuting connections. The ones shown are those with bigger fluxes. \textbf{b)} County population distribution. \textbf{c)} Commuting fluxes distribution. \label{distris_comm}}
\end{figure}

The commuting data is taken from the US census of year $2001$. It provides the population of each county and the number of individuals $N_{ij}$ living in county $i$ and working in county $j$, where $i,j$ is a couple of counties with a non-vanishing flux of commuters. The data contains $3117$ counties or county-equivalent regions with an average population of $89585$ and a standard deviation of $292405$. The whole distribution is shown in Figure~\ref{distris_comm} bottom left, where one can see the broad nature of it. There are $162131$ commuting connections between different counties with a mean flux of $10854$ individuals and a standard deviation of $15584$. The whole distribution is shown in Figure~\ref{distris_comm} bottom right. Note that this forms a directed weighted network, with $3117$ nodes corresponding to the counties and $162131$ directed and weighted edges encoding the number of people living in one county and working in another. If we add one weighted self-loop per county counting how many 
individuals live and work at each county, we have embedded all population and commuting data in a network structure.

\begin{figure}[H]
\begin{center}
 \includegraphics[width=\textwidth]{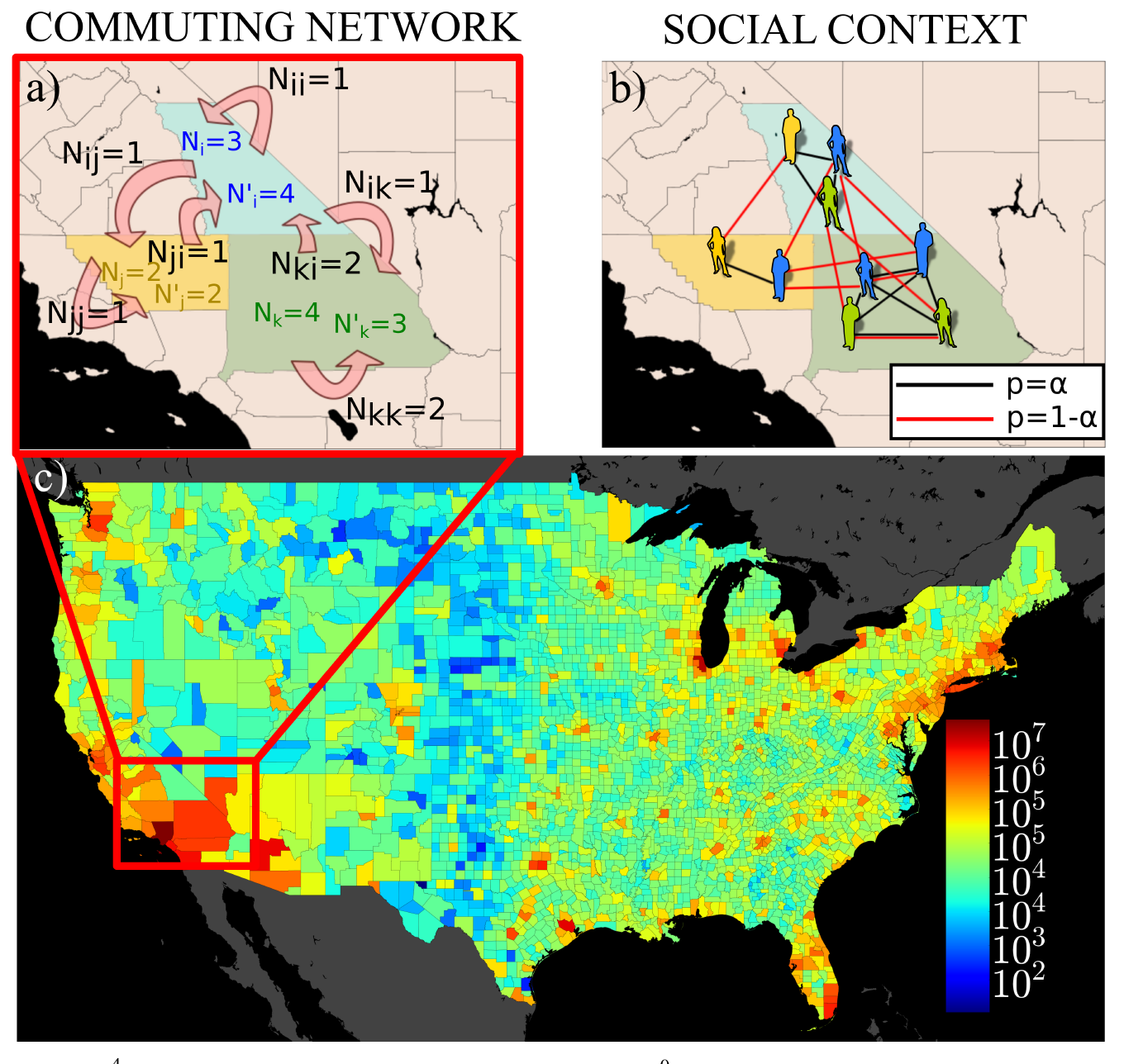}
 \caption{\textbf{Recurrent mobility and population heterogeneities.} \textbf{a)} Schematic representation of the commuting network obtained from census data. \textbf{b)} Schematic representation of the different agent interactions. The home county interactions (black edges) and work county interactions (red edges) occur with different probabilities ($\alpha$ and $1-\alpha$ respectively). The agents are placed at their home counties and colored by their work counties. \textbf{c)} Map of the populations by county in the 2001 census. The color scale is logarithmic because there are populations ranging from around a hundred to several million individuals.  \label{model}}
 \end{center}
\end{figure}

In Fig.~\ref{model} a schematic representation of how to construct the social context of the individuals starting from the commuting data is shown. There is also a map showing the spatial distribution of populations.

\newpage

\section{Election data}
We use the results of the presidential elections from 1980 to 2012 aggregated by counties. The analysis of the data unravels statistical characteristics of US elections.
\subsection{National vote}
\begin{figure}[H]
 \centering
 \includegraphics[width=\textwidth]{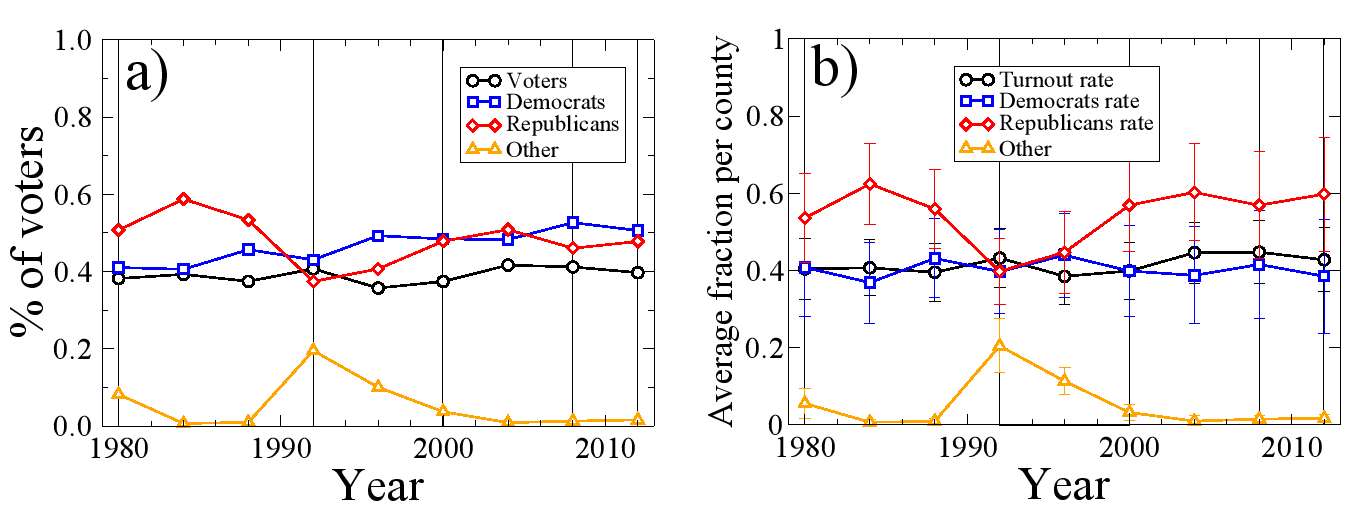}
 \caption{ \textbf{National election results.} The colors of the background indicate the president's party (red for republican and blue for democrat). \textbf{a)} Global trends for the absolute values of different quantities such as population, turnout, votes for democrats, republicans and other. \textbf{b)} Global trends for the percentages of different quantities such as turnout, fractions of votes for democrats, republicans and other. The dots are the average over all counties for different years and the bars represent the standard deviation of those averages.\label{global}}
\end{figure}

In this section we show global features of the US presidential electionsfrom 1980 to 2012. Namely we show the turnout, votes for democrats, republicans and others in Fig. \ref{global}a. In Fig. \ref{global}b we show the evolution of the global shares associated to turnout and votes for the different parties. The shares are computed county by county and then we extract the average and its standard deviation.

\newpage

\subsection{Per county vote and spatial correlations}
\begin{figure}[H]
 \centering
 \includegraphics[width=\textwidth]{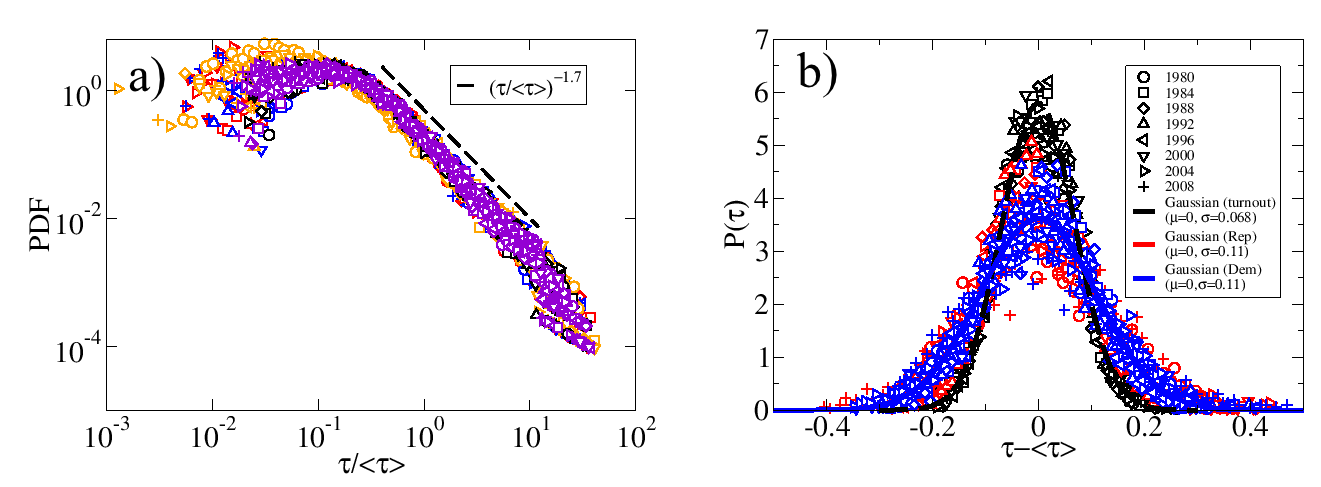}
 \caption{\textbf{Per county distributions.} \textbf{a)} Distributions of the absolute values of population (violet), turnout (black), votes for democrats (red), votes for republicans (blue) and votes for other (orange). The distributions are rescaled in such a way that they all have average equal to 1. All of them collapse to a single curve with a power-law decay with exponent $1.7$. The different symbols refer to different years. \textbf{b)} Turnout fraction, democrat and republican vote fraction distributions for all elections as a function of the fraction minus the average . They follow a Gaussian distribution. It seems that both republican and democrat follow the same distribution, which is wider than the one that is followed by the turnout fractions.\label{colapse_distris_abs}}
\end{figure}

In Fig. \ref{colapse_distris_abs} a) we plot the distributions of turnout, population and votes for the different parties for all years in the dataset ($1980$--$2012$), properly rescaled to have mean equal to $1$. In Fig. \ref{colapse_distris_abs} b) we plot the distributions of turnout fraction and vote shares for the democrats and republicans, once discounted the average for each year.

\begin{figure}[H]
 \centering
 \includegraphics[width=\textwidth]{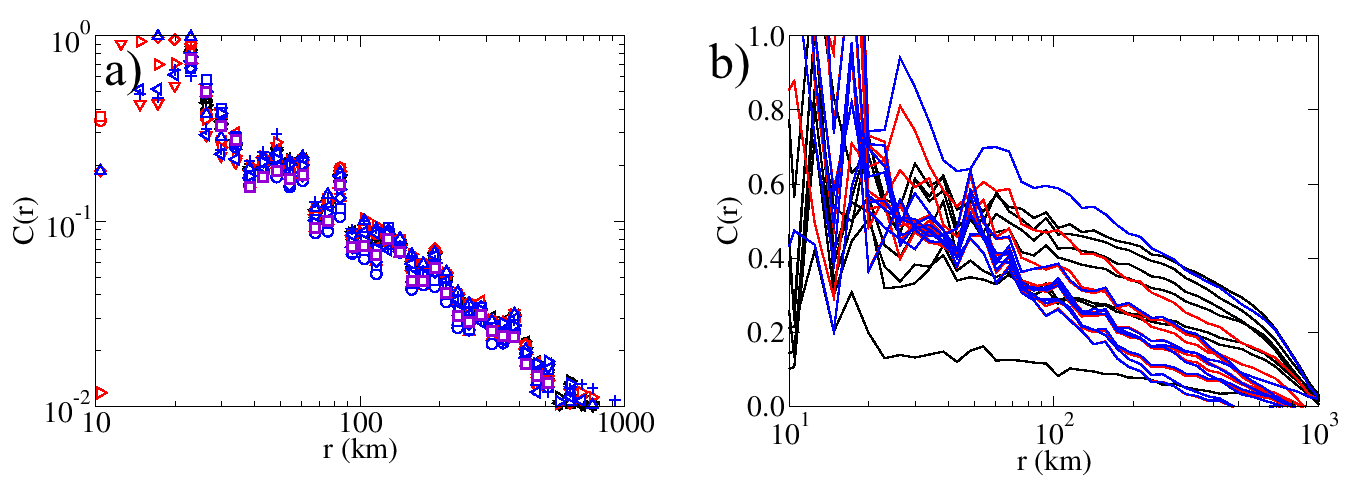}
 \caption{\textbf{Spatial correlations.} \textbf{a)} Correlations between absolute values show a power-law decay with exponent around $1.2$. The data in this figure is for turnout (black), votes for democrats (blue), republicans (red) for all years in the dataset and population (violet). Different symbols refer to different years. \textbf{b)} Correlations between fractions of values show a logarithmic decay.}
\end{figure}\begin{figure}[H]
 \centering
 \includegraphics[width=\textwidth]{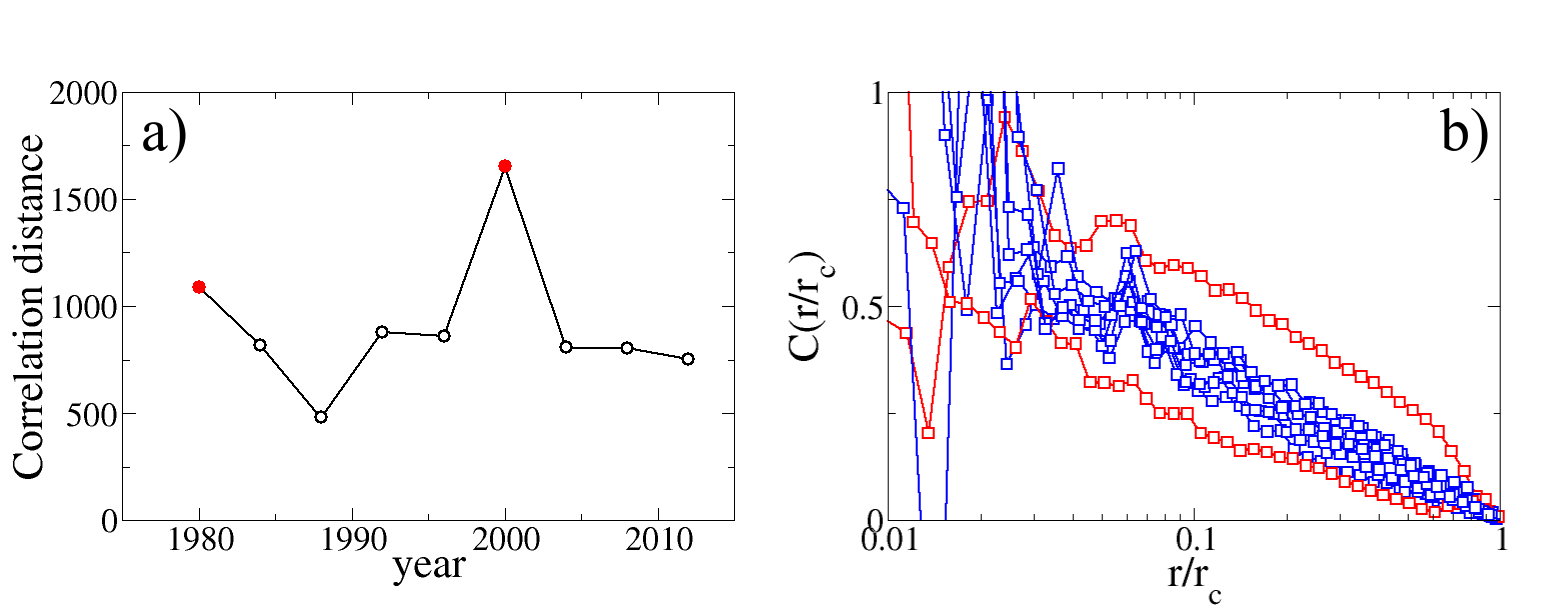}
 \caption{\textbf{Correlation distance.} \textbf{a)} Correlation distance for the vote-share correlations as a function of the year. The correlation distance is defined as the distance at which the correlation first crosses $0$. \textbf{b)} Correlations on a distance axis rescaled by the correlation distance of each year. Therefore all curves cross $0$ at normalized distance $1$. All the curves collapse nicely except for two of them, corresponding to years $1980$ and $2000$, which are the ones marked in red in a). }
\end{figure}

\newpage

\section{Aggregation across geographical scales}
\begin{figure}[H]
 \centering
 \includegraphics[width=\textwidth]{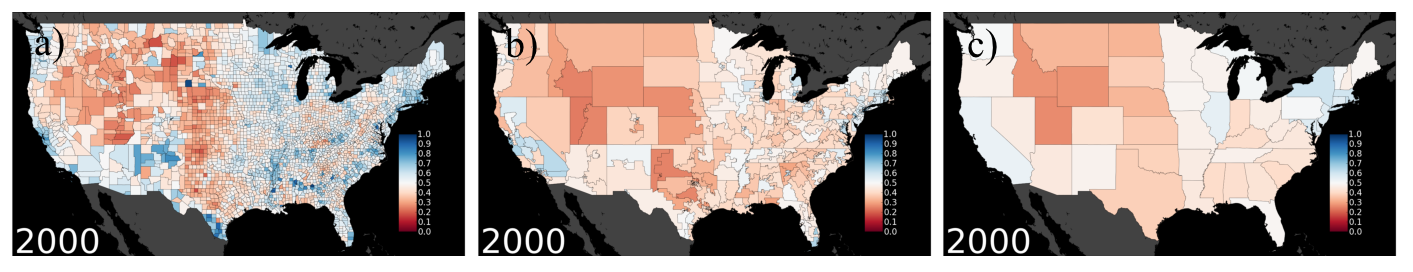}
 \caption{Aggregation to bigger geographical areas of the real data of year $2000$. Spatial configuration of democrat vote shares per county (a), per congressional district (b) and per state (c). The boundary files for counties, congressional districts and states where taken from the census web page~\cite{census}.\label{maps_renorm}}
\end{figure}

\begin{figure}[H]
 \centering
 \includegraphics[width=\textwidth]{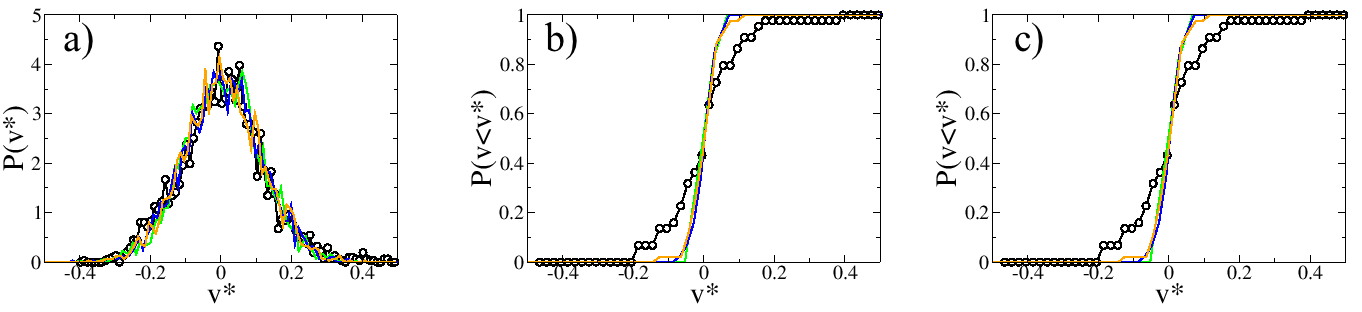}
 \caption{\textbf{Uncorrelated aggregation.} Comparison of the aggregation to bigger geographical areas of the real data of year $2000$ (other years look very similar) and randomized data. Randomized data does not aggregate in the same way. \textbf{a)} County vote share distribution. The black circles show the democrat data of year $2000$, while the other curves are just random assignations of vote shares following the same distribution. \textbf{b)} Aggregation to show the cumulative distribution of congressional districts vote shares. The randomized data do not aggregate as the real data. \textbf{c)} Aggregation to show the cumulative distribution of state vote shares. The randomized data do not aggregate as the real data.\label{renorm_random}}
\end{figure}

The way in which election data aggregate can be seen in the maps of Fig.\ref{maps_renorm}.
Here we show that the result of aggregating for bigger geographical areas than counties, \textit{i.e.}, congressional districts or states, is strongly dependent on the spatial configuration of the election results. For doing so we compare the result of this aggregation for real data from year $2000$ and the result from the aggregation procedure of a random configuration of county shares that follows the same distribution as the one displayed by the data. This comparison can be seen in Figure~\ref{renorm_random}.
\section{Data vs. model predictions}
\begin{figure}[H]
 \centering
 \includegraphics[width=\textwidth]{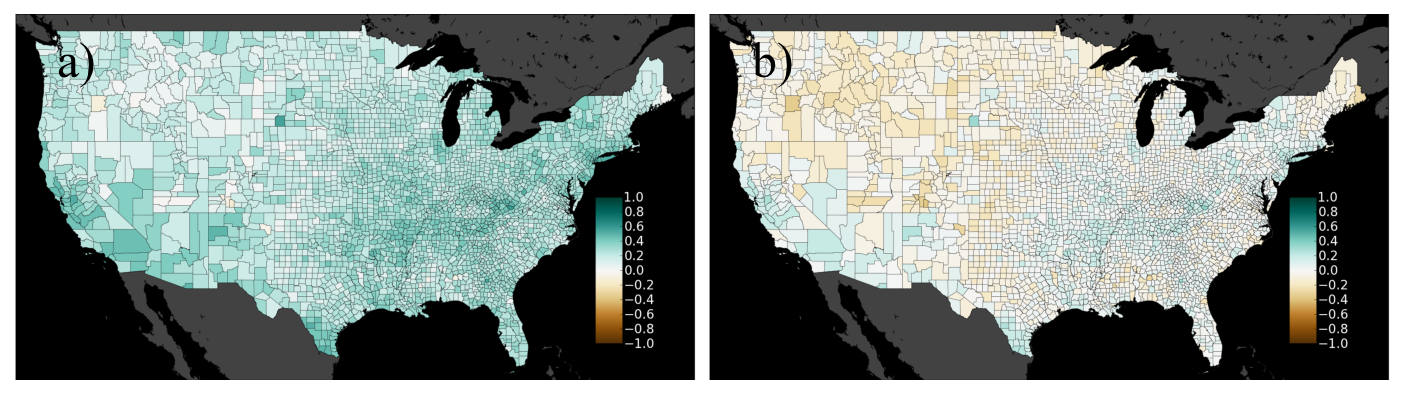}
 \caption{\textbf{Difference between data and model prediction.} Maps showing the difference between real data and model after 12 years. The model is evolved for $12$ years, starting from the initial condition from the data of year $2012$, with parameters $\alpha=1/2$ and $D=0.02$. Then the results of the model are compared to the electoral results of year $2012$. \textbf{a)} Direct substraction of data minus model. \textbf{b)} For this we first substract the national average both from data and model results and then do the substraction of data minus model. This image shows that all values are very near to zero, thus being model and data in good agreement. The point here is that the model describes the fluctuations in election data and does not account for the real average value of the vote shares.}
\end{figure}

\newpage

\section{Dependence of the results on $\alpha$}

\begin{figure}[H]
 \centering
 \includegraphics[width=0.48\textwidth]{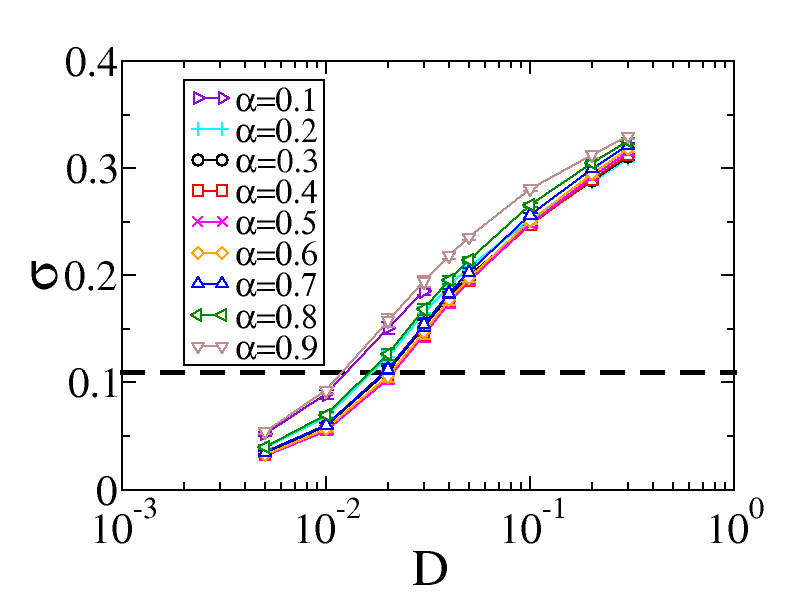}
 \includegraphics[width=0.48\textwidth]{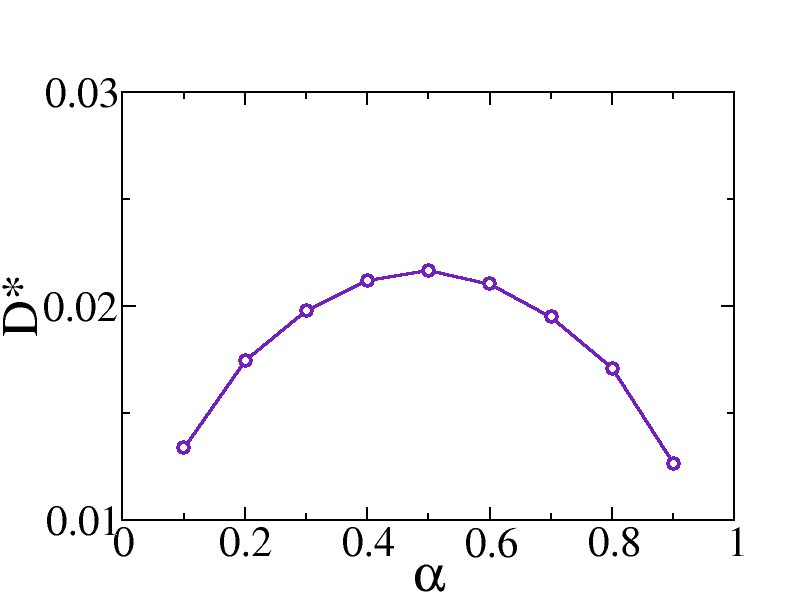}\\
 \includegraphics[width=0.48\textwidth]{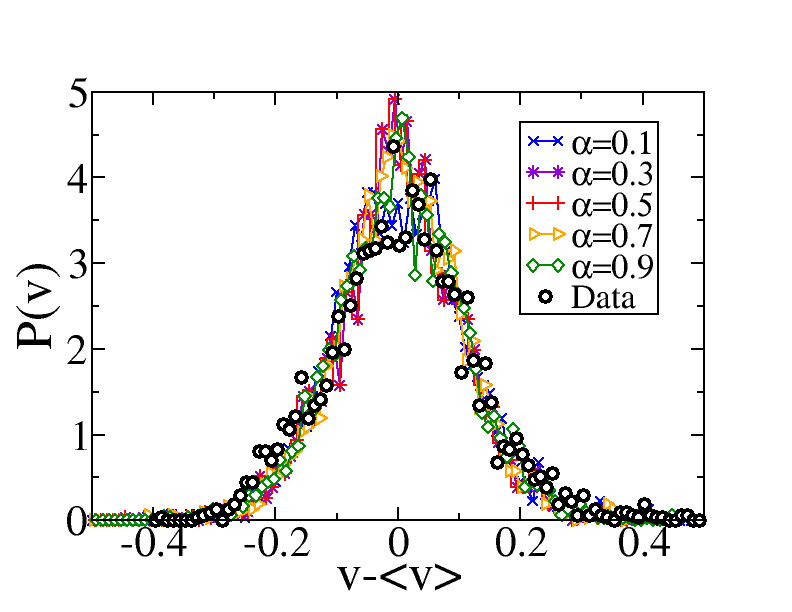}
 \includegraphics[width=0.48\textwidth]{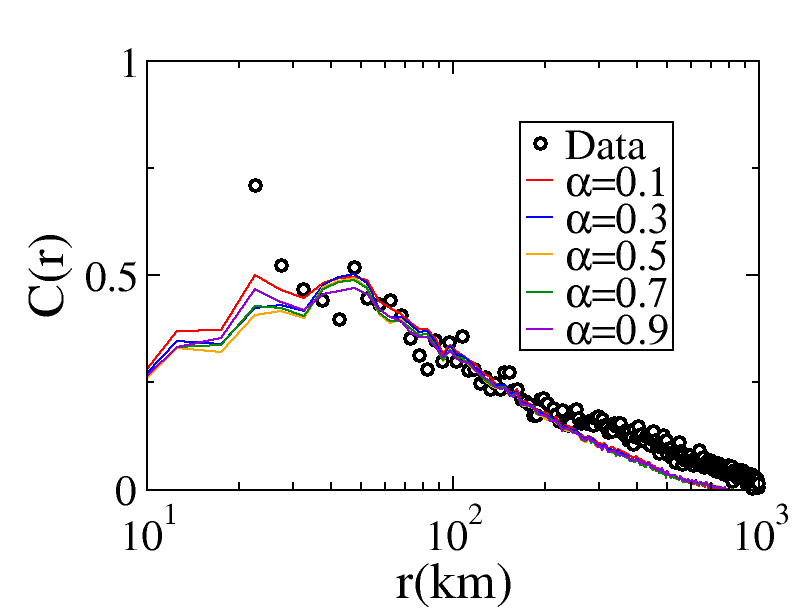}
 \caption{\textbf{Exploration of $\alpha$.} \textbf{Top left:} Calibration curves for different values of $\alpha$ on the full commuting network. The curves show the standard deviation of the vote-share distribution after $10000$ Monte Carlo steps. \textbf{Top right:} Value of the noise intensity $D^*$ that recovers the empirical value of the standard deviation of the vote-share distribution. \textbf{Bottom left:} Vote-share distributions after $10000$ Monte Carlo steps for different values of the parameter $\alpha$ at the calibrated noise intensity $D^*$. \textbf{Bottom right:} Spatial correlations after $10000$ Monte Carlo steps for different values of the parameter $\alpha$ at the calibrated noise intensity $D^*$.\label{alphas}}
\end{figure}

Here we show that the results shown in the main text do not depend crucially on parameter $\alpha$, unless if $\alpha=0$ or $1$. In that case the system consists of disconnected patches and thus there is no spatial diffusion. We exclude these cases from the analysis below. For the other cases one can intuitively see from the dynamical equations that a variation in $\alpha$ will change the timescales of the model and the values of the noise intensity $D$ to recover the empirical standard deviation of vote-shares. In Fig.~\ref{alphas} we show the calibration of the model on the full commuting network for different values of $\alpha$. Although the value at which the model is calibrated depends on $\alpha$, the properties of the model at that point remain as in the case of $\alpha=1/2$, i.e., the vote-share distributions remain stationary and the spatial correlations fall logarithmically in space.

\newpage

\section{Derivation of the model equations}
First let us derive Equation (2) of the main text. The rates of the model are
\begin{align}
 r_{ij}^+(V_{ij}\rightarrow V_{ij}+1)&=(N_{ij}-V_{ij})\left[\alpha\frac{V_i}{N_i}+(1-\alpha)\frac{V'_j}{N'_j}\right]+N_{ij}\frac{D}{2}\eta_{ij}^+(t),\nonumber\\
 r_{ij}^-(V_{ij}\rightarrow V_{ij}-1)&=V_{ij}\left[\alpha\frac{N_i-V_i}{N_i}+(1-\alpha)\frac{N'_j-V'_j}{N'_j}\right]+N_{ij}\frac{D}{2}\eta_{ij}^-(t).
\end{align}
For a review on models with stochastic rates see Ref.~\cite{maxi_external_noise}. Given these rates one can write down the master equation for the probability $P(\mathcal{V};t)$ of having $V_{11}$ agents with state $+1$ in subpopulation $11$, $V_{12}$ agents with state $+1$ in subpopulation $12$, and so on at time $t$. We take the notation $\mathcal{V}=\{V_{11},V_{12},\dots,V_{ij},\dots,V_{nn}\}$ and $\mathcal{V}_{ij}^{\pm}$ is equal to $\mathcal{V}$ except for $V_{ij}$, wich is replaced by $V_{ij}\pm1$. Then the master equation is
\begin{equation}
 \frac{\partial P(\mathcal{V};t)}{\partial t} = \sum_{i,j} \left[r_{ij}^+(\mathcal{V}_{ij}^-) \, P(\mathcal{V}_{ij}^-;t)+r_{ij}^-(\mathcal{V}_{ij}^+) \,P(\mathcal{V}_{ij}^+;t)-\left(r_{ij}^+(\mathcal{V})+r_{ij}^-(\mathcal{V})\right)\, P(\mathcal{V};t)\right].
\end{equation}
By standar methods one can find a Fokker Planck equation that approximates this master equation,

\begin{equation}
\begin{split}
 \frac{\partial P(\mathcal{V};t)}{\partial t} = \sum_{i,j} \Bigg\{ &-\frac{\partial}{\partial V_{ij}}\left[\left(r_{ij}^+(\mathcal{V})-r_{ij}^-(\mathcal{V})\right)P(\mathcal{V};t)\right]\\&+\frac{\partial^2}{\partial V_{ij}^2}\left[\frac{1}{2}\left(r_{ij}^+(\mathcal{V})+r_{ij}^-(\mathcal{V})\right)P(\mathcal{V};t)\right]\Bigg\}.\nonumber
\end{split}
\end{equation}

We can translate the Fokker-Planck equation into a Langevin equation, which will describe the dynamics of the numbers of voters with state $+1$ in each subpopulation, $V_{ij}$. Here we already show this equation for the densities $v_{ij}=V_{ij}/N_{ij}$
\begin{align}\label{langevin_eq}
 &\frac{dv_{ij}}{dt}=\alpha\sum_l\left(\frac{N_{il}}{N_i}-\delta_{jl}\right) v_{il}+(1-\alpha)\sum_l\left(\frac{N_{lj}}{N'_j}-\delta_{li}\right) v_{lj}+ D \eta_{ij}(t)\\&+\frac{1}{\sqrt{N_{ij}}}\sqrt{(1-2v_{ij})\left(\alpha\frac{\sum_lN_{il}v_{il}}{N_i} +(1-\alpha)\frac{\sum_lN_{lj}v_{lj}}{N'_j}\right)+v_{ij}+\frac{D}{2}\eta_{ij}^{'}(t)}\eta_{ij}^*(t).\nonumber
\end{align}
Note also that in the right side of Eq.(\ref{langevin_eq}) all the terms are of order $1$ (densities) except for the last term, which accounts for the variability of a single realization of the stochastic process and is of order $1/\sqrt{N_{ij}}$. Given the sizes of the subpopulations $N_{ij}$ it is reasonable to disregard this term. The error will be of more importance for smaller subpopulations.\\
The ensemble average of Eq.(\ref{langevin_eq}) reveals a Laplacian equation
\begin{equation}\label{ensemble_eq}
 \frac{d\langle v_{ij}\rangle}{dt}=\alpha\sum_l\left(\frac{N_{il}}{N_i}-\delta_{jl}\right) \langle v_{il}\rangle +(1-\alpha)\sum_l\left(\frac{N_{lj}}{N'_j}-\delta_{li}\right) \langle v_{lj}\rangle.
\end{equation}
This dynamics conserves the number of voters with state $+1$, $\sum_{i,j}N_{ij}\langle v_{ij}\rangle$ and reaches asymptotically a homogeneous configuration with $v_{kl}=\frac{1}{N}\sum_{ij}N_{ij}\langle v_{ij}\rangle$ for all subpopulations $kl$.\\

\newpage

\section{Geographical component of the coupling between counties}

\begin{figure}[H]
 \centering
 \includegraphics[width=0.49\textwidth]{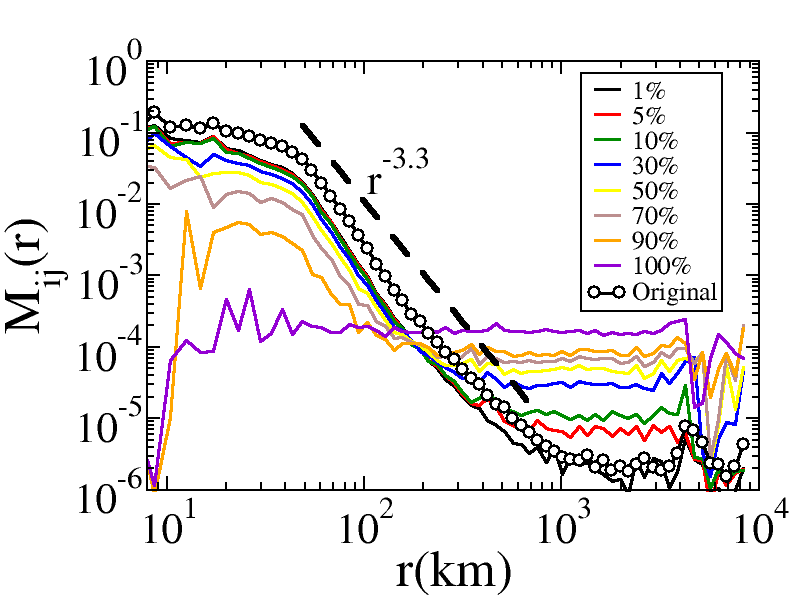}
 \includegraphics[width=0.49\textwidth]{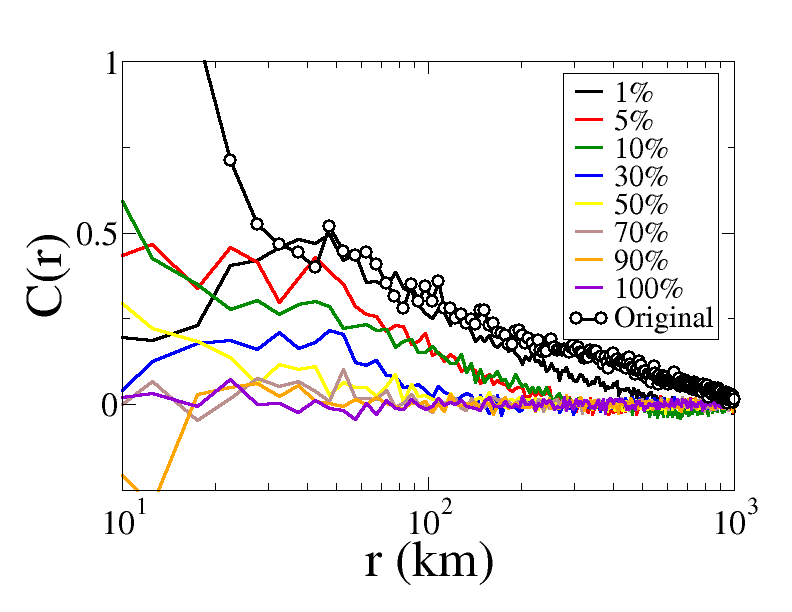}
 \caption{\textbf{Left:} Average coupling as a function of distance for the fast mixing approximation. The black empty circles correspond to the average coupling as a function of distance given the original commuting network. The other curves correspond to the coupling as a function of distance for different partial randomizations of the commuting network. \textbf{Right:} Asymptotic correlation function for the different cases. The color code is the same as in the left panel.}\label{contact_kernel}
\end{figure}

From Eq.(\ref{ensemble_eq}) one can derive an approximation, which we call the fast mixing approximation. It considers that the opinions are infinitely fast mixed at home, thus densities of voters with state $+1$ who live in the same location are all the same, \textit{i.e.}, $\langle v_{ij}\rangle=\langle v_{il}\rangle=\langle v_{i}\rangle$ for any $i$,$j$ and $l$. After multiplying the deterministic part of Eq.(3) by $N_{ij}$, summing over $j$ and dividing by $N_i$,it takes the form
\begin{equation}
 \frac{d\langle v_i\rangle}{dt}=(1-\alpha)\sum_j\left[\sum_l\frac{N_{jl}N_{il}}{N_iN_l^{'}}-\delta_{ij}\right]\langle v_j\rangle=(1-\alpha)\sum_j M_{ij} \langle v_j\rangle.
\end{equation}
This equation keeps the Laplacian nature of the dynamics and represents a voter model on a directed weighted network with self-loops. One can now look at the average coupling strength as a function of distance by averaging entries of the matrix $M_{ij}$ which couple locations $i$ and $j$ separated a distance $r$.
When using the original commuting network in the US the average coupling $M_{ij}(r)$ decays with distance but it is still not clear how the decay must be in order to recover the usual diffusion characteristics in 2d. Nevertheless it seems that for the commuting network from the data, the coupling as a function of distance decays fast enough to recover the logarithmic correlations typical of a diffusive process in two dimensions.\\
We randomized the network by randomly rewiring links with a certain probability $p$. As can be seen in Fig.~\ref{contact_kernel} the coupling as a function of distance gets flat when the network is fully randomized and the corresponding correlations for the calibrated model are zero. Nevertheless if the commuting network is not fully randomized, the coupling still presents a decay that ends in a plateau. From the correlations for those networks it is clear that the logarithmic decay still appears, but gets to zero correlation much closer to zero as the network is more and more random.

\begin{thebibliography}{10}
\bibitem{footnote3} In the derivation of the master equation we set the differential of time to be $\delta t=1/N$. This leads to Eq.(3) which is written in arbitrary units, as indicated in the revised manuscript. The choice of $\delta t=1/N$ makes the arbitrary time unit identifiable to the MC step. Thus in the case of the full commuting network, where $2.5$ MC steps correspond to a year, the arbitrary time unit in the Langevin equation corresponds to $0.4$years. In order to have the Langevin equation written such that the time unit is one year, one should rescale time by a factor $a=2.5$ and both the deterministic part of the equation and the noise term should be properly rescaled.
\end{thebibliography}
\end{document}